\documentclass[twocolumn,prb,amsmath,amssymb,superscriptaddress]{revtex4}

\usepackage{graphicx}
\usepackage{color}

\def\horparallel{ \lower.5ex\hbox{ \includegraphics[width=2ex]{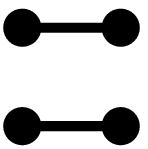}}\,\, }

\def\vertparallel{ \lower.5ex\hbox{ \includegraphics[width=2ex]{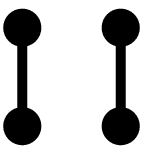}}\,\, }

\begin{document}

\title{Continuous interpolation between the fully frustrated Ising
and quantum dimer models}

\author{Dmitri A.~Ivanov}
\affiliation{Institute of Theoretical Physics,
Ecole Polytechnique F\'ed\'erale de Lausanne (EPFL),
CH-1015 Lausanne, Switzerland}

\author{S.~E. Korshunov}
\affiliation{L.~D.~Landau Institute for Theoretical Physics RAS, 142432,
Chernogolovka, Russia}

\date{June 16, 2011}


\begin{abstract}
We propose a new quantum model interpolating between the fully frustrated
{spin-1/2} Ising model in a transverse field and a dimer model. This model
contains a resonating-valence-bond phase, including a line with an exactly
solvable ground state of the Rokhsar--Kivelson type. We discuss the phase
diagrams of this model on the square and triangular (in terms of the dimer
representation) lattices.
\end{abstract}

\maketitle

\section{Introduction}

The concept of the resonating-valence-bond (RVB) phase \cite{Lhuillier2}
has remained for many years one of the most fascinating topics in the
physics of strongly correlated systems. This phase was originally proposed
for frustrated spin systems \cite{Fazekas} and for high-temperature
superconductors, \cite{PWA} but to date no confirmed examples of physical
frustrated magnets with the RVB phase have been found and the relevance of
RVB physics to high-temperature superconductivity remains at the level of
conjecture. However, many model systems
have been proposed to exhibit the RVB phase, \cite{Balents-Fisher-Girvin,%
Senthil-Motrunich,Ioffe-Feigelman,Seidel} including quantum dimer models,
\cite{MS01prl,Misguich-Serban-Pasquier} where this phase is most easily
accessible. Interest in the model studies of the RVB phase has been
further stimulated by experimental \cite{kagome-experiment} and
theoretical \cite{kagome-numerics} works on the spin-1/2 Heisenberg model
on the kagome lattice, which suggested a spin-liquid phase. More recently,
it has been suggested that this kagome-lattice model is closely related to
a certain class of quantum dimer models. \cite{kagome-poilblanc} In view
of these developments, model studies of phase diagrams of various systems
involving the RVB phase may be useful for further search and
identification of this phase. In the present work, we propose a new class
of models that realize an interpolation between quantum dimer models
(QDMs) and fully frustrated spin-1/2 Ising models in a transverse field
(FFIMs) with an exactly solvable RVB ground state.

There is a well-known correspondence between QDMs and FFIMs.
\cite{MSCh,MS01prb} When these two models are formulated on dual lattices
(so that the sites of the Ising lattice correspond to the plaquettes of
the dimer lattice and vice versa), frustrated and unfrustrated bonds of
the Ising model can be put into correspondence with the presence and
absence of a dimer, respectively. Therefore, in the limit of a strong
Ising coupling (compared to the transverse field), the number of
frustrated bonds at each Ising plaquette should be odd and minimal, which
implies the QDM constraint of fully packed dimers.

An exact mapping between the two models exists, however, only in the
above-mentioned limit of the strong Ising coupling. In the well-studied
examples of square and triangular latices, this limit belongs to the
crystallized phase (dimers or Ising spins order and break the
translational symmetry of the lattice). \cite{MS01prl,Sachdev89,Syl06} On
the other hand, in the opposite limit of a strong transverse field FFIM is
in the disordered phase quite similar to the RVB phase realizable in QDMs
on nonbipartite lattices. The characteristic features of the RVB phase are
unbroken translational invariance (on lattices with a single site per unit
cell), exponential decay of correlations, topological order on multiply
connected domains, and Z$_2$ vortices (visons) as elementary excitations.
\cite{Read-Chakraborty,KRS,Senthil-Fisher}

Although a conjecture was made  that the disordered phase of the FFIM is
continuously connected to the RVB phase of the
QDM,\cite{MS01prl,Misguich-Mila} this never was demonstrated explicitly.
In the present work we provide a justification for such an identification
of the disordered FFIM phase and the RVB dimer phase by constructing a
more general model which realizes a continuous interpolation between
FFIMs and QDMs. This model is exactly solvable (in terms of its ground
state) along a line belonging to the RVB phase and connecting the
high-field limit of the FFIM to the Rokhsar--Kivelson point of the QDM.

The paper is organized as follows. In Sec. \ref{sec:dimer-spin}, we
briefly review the QDM and the FFIM and their RVB phases. In Sec.
\ref{sec:interpolation}, we construct the interpolating model connecting
the QDM and the FFIM on mutually dual lattices. In Secs. \ref{sec:square}
and \ref{sec:triangular}, we analyze in more detail the phase diagrams of
this interpolating model on the square and triangular (in terms of the
dimer representation) lattices, respectively. Finally, in Sec.
\ref{sec:conclusion}, we summarize and discuss our findings.

\section{RVB states in dimer and spin models}
\label{sec:dimer-spin}

\subsection{Quantum dimer model}

The simplest QDM is the so-called Rokhsar--Kivelson (RK) dimer model:
\cite{RK}
\begin{equation}
H_{\rm RK}= \sum \Big( -t
\left| \horparallel  \right\rangle\left\langle \vertparallel \right|
+ v \left| \horparallel  \right\rangle\left\langle \horparallel \right|
\, \Big)
\label{rk-hamiltonian}
\end{equation}
where the sum is taken over all tetragonal plaquettes of the lattice. At
$v=t$, both on square and triangular lattices, the ground state of the RK
model (or, possibly, one of the ground states, depending on the topology
of the lattice cluster) is given exactly by the sum of all possible dimer
configurations with equal amplitudes (the RK state). \cite{RK}

On the {\em triangular lattice}, the RK state is an exemplary RVB
state: it has a finite gap and exponentially decaying correlations,
topological degeneracy and visons as excitations.
\cite{MS01prl,FMS,IIF,Ivanov04} It however lies at
the phase boundary: at $v>t$, the model crystallizes via a first-order
phase transition into the so-called staggered (or nonflippable) state. On
the other side, at $v<t$, there is a finite window of RVB states.
According to numerical studies, \cite{MS01prl,Ralko} this window
extends to approximately $v/t=0.83(2)$.

On the {\em square lattice}, at $v=t$, the RK state is critical: it has
gapless excitations and power-law correlations.
On both sides of the point $v=t$, the system crystallizes as soon as one
deviates from this point. \cite{Syl06,LChR,RPM,FHMOS}

The difference between the square and triangular lattices results from one
of them being bipartite while the other is not. There is a vast literature
on the phase diagram of the RK dimer models on both lattices and on the
properties of the RVB states (for a review, see Ref.\
\onlinecite{moessner-raman}).

\vspace*{-5mm}

\subsection{Fully frustrated Ising model}

The Hamiltonian of the FFIM reads
\begin{equation}
H_{\rm FFIM}
= -J \sum_{\{ij\}} M_{ij} \sigma^z_i \sigma^z_j - \Gamma \sum_i \sigma^x_i
\label{ffim-hamiltonian}
\end{equation}
where the coefficients $M_{ij} = \pm 1$ are chosen in such a way that
their product over any plaquette of the Ising lattice is negative (with
both parameters $J$ and $\Gamma$ assumed to be positive). \cite{MS01prb}
Here and below we denote the positions of Ising spins by Latin indices
$i$, $j$, etc., whereas the first sum in Eq.~(\ref{ffim-hamiltonian}) is
taken over all pairs of nearest neighbors $ij$.

In this model, the RVB state is obviously realized in the limit of strong
transverse magnetic field $\Gamma \gg J$. In this limit, the ground state
has all spins almost fully polarized along the field, their transverse
components being disordered. Since each spin aligned in the $x$ direction
is a linear combination of spins $\sigma^z=\pm 1$ with equal amplitudes,
the resulting ground state contains all sets of $z$ projections of spins
with almost equal amplitudes,\cite{MS01prl} which resembles the RVB state
in the RK model. This state is translationally invariant, has only local
correlations, and has an excitation gap of order $\Gamma$. The Ising spins
in the FFIM model correspond to the vison operators in the QDM.
\cite{Misguich-Serban-Pasquier,MS03}

\vspace*{-2mm}
\subsection{Relation between the two models}

A rigorous mapping between the FFIM and the QDM exists only in the
zero-field limit ($\Gamma/J \to 0$) of the FFIM when it becomes equivalent
to the RK model with $v/t = 0$. \cite{MSCh} This mapping involves FFIMs
and QDMs on mutually dual lattices (i.e., plaquettes of one lattice
correspond to sites of the other). While the construction of the
interpolating model developed in the next section is generally applicable
to any dual pair of lattices, we will further illustrate it with specific
examples of the triangular and square QDM lattices (which corresponds to
the hexagonal and square FFIM lattices, respectively). On both of these
lattices, the point at which there exists a rigorous mapping between the
FFIM and the QDM corresponds to crystallized phases.
\cite{MS01prb,Sachdev89,Syl06}

Below we explicitly construct a model which realizes a continuous
interpolation between the FFIM and the QDM on mutually dual lattices and
is exactly solvable along the line connecting the FFIM at $\Gamma/J \to
\infty$ (the limit of noninteracting spins) to the RK model at
\makebox{$v=t$}. Along this line the system belongs to the RVB phase.

\section{Interpolating model}
\label{sec:interpolation}

\subsection{Construction of an RK-type model}
\label{exactly-solvable-model}

For the interpolating model, we use the same Hilbert space as in the FFIM.
It will be convenient to introduce the basis ${\cal Z}$ defined in terms
of projections of spins on axis $z$. Then we first postulate the ground
state of the interpolating model parametrized by a ``chemical potential''
$\mu$:
\begin{equation}
\Psi_\mu = \sum_{|c\rangle \in {\cal Z}}
\exp\left[ \frac{\mu}{2} \sum_{\{ij\}} 
M_{ij} \sigma^z_i \sigma^z_j
\right]\, |c\rangle \, ,
\label{gs-interpolation}
\end{equation}
where the sum is taken over all Ising-spin configurations $| c\rangle$
from the basis ${\cal Z}$. To avoid confusion, the lattice at whose sites
the variables $\sigma_j$ are defined is called below the FFIM lattice and
the lattice dual to it the QDM lattice. The sites of the latter are
denoted by Greek letters. Note that the construction described in this
section is rather general and does not require the lattices to be
periodic.

In terms of the dimer representation, the quantity
\begin{equation}
\tau_{ij} = M_{ij} \sigma^z_i \sigma^z_j\,
\end{equation}
describes the number of dimers $n_{\alpha\beta}= (1-\tau_{ij})/2$
on the bond $\alpha\beta$ of the QDM lattice which crosses the bond $ij$
of the FFIM lattice (i.e., $\tau_{ij}$ equal to plus or minus one
corresponds to the absence or presence of a dimer, respectively). Because
of the frustration of the parameters $M_{ij}$, each site of the dimer
model lattice must belong to an odd number of dimers. In the limit $\mu\to
\infty$ only configurations with exactly one dimer per site survive, thus
the ground state (\ref{gs-interpolation}) continuously interpolates
between the state which is fully polarized in the $x$ direction (the
ground state of FFIM in the limit $\Gamma/J\to \infty$) at $\mu=0$ and the
RK state (the ground state of RK dimer model at $v=t$) at $\mu \to
\infty$.

At the next step, we construct the Hamiltonian whose ground state is given
by (\ref{gs-interpolation}). This is performed in the ``supersymmetric''
way suggested by Henley \cite{Henley}; namely, the Hamiltonian is assumed
to have a form of a quadratic sum
\begin{equation}
H=\sum_\alpha Q^\dagger_A Q_A\;, \label{SUSY-general}
\end{equation}
where the operators $Q_A$ are such that they annihilate the chosen ground
state. We call such a class of models the {\it RK-type} models.

The decomposition (\ref{SUSY-general}) for the standard RK model
(\ref{rk-hamiltonian}) is well known, with the operators $Q_A$ removing
two dimers on one tetragonal plaquette $\alpha$ in two possible ways with
opposite signs. \cite{moessner-raman} In a shorthand notation, this
operator may be written as
\begin{equation}
Q_A = \left\langle \vertparallel \right|_A - \left\langle \horparallel
\right|_A\, . \label{rk-Q}
\end{equation}
Note that this operator does not acts in the physical Hilbert space but
maps physical configurations of dimers onto configurations in some
auxiliary space with one plaquette removed.

The above construction can be directly generalized to Ising-type systems
to annihilate the ground state (\ref{gs-interpolation}). The simplest
choice of operators $Q_A$ involves removing one Ising spin,
\begin{equation}
Q_i = \Big(\left\langle \uparrow \right| -
\left\langle \downarrow \right| \Big)_i \,
\exp\left[ - \frac{\mu}{2} \sum_{j={\rm n.n.}(i)} \tau_{ij} \right] \, ,
\label{type1-Q}
\end{equation}
where the sum over $j$ is taken over the nearest neighbors of the site $i$
denoted ${\rm n.n.}(i)$.

After some simple algebra, the Hamiltonian (\ref{SUSY-general})
with operators $Q_i$ given by (\ref{type1-Q})
can be rewritten as
\begin{equation}                    \label{model1-hamiltonian}
H = \sum_i \exp\left[ - \mu \sum_{j={\rm n.n.}(i)} \tau_{ij} \right]
-\sum_i \sigma^x_i \, ,
\end{equation}
where the sums over $i$ are taken over all sites of the FFIM lattice. The
sum over $j$ for each $i$ is taken over the nearest neighbors of $i$.

In terms of the dimer representation the first term in
Eq.~(\ref{model1-hamiltonian}) can be rewritten as $\sum_{i} \exp
\left[-{\mu}(m_i-2n_i)\right]$. Here $m_i$ is the number of nearest
neighbors of site $i$ on the FFIM lattice (i.e., the number of bonds
belonging to ${\cal P}_i$, the corresponding plaquette of the QDM lattice)
and \makebox{$n_i=\sum_{\{\alpha\beta\}\in{\cal P}_i} n_{\alpha\beta}$} is
the total number of dimers on ${\cal P}_i$. In the case of a simple
periodic lattice $m_i={\rm const}$. At the same time, operator
$\sigma^x_i$ corresponds to the inversion of dimer occupation numbers
($n_{\alpha\beta}\mapsto 1-n_{\alpha\beta}$) on all bonds belonging to
${\cal P}_i$.

The resulting Hamiltonian (\ref{model1-hamiltonian}) is an RK-type model
which continuously interpolates between the system of non-interacting
spins in a uniform magnetic field [in other terms, the \makebox{$\Gamma/J
\to \infty$} limit of the FFIM (\ref{ffim-hamiltonian})] at $\mu\to 0$ and
a dimer model with the RK ground state at $\mu\to\infty$. More precisely,
in the $\mu\to\infty$ limit, the model (\ref{model1-hamiltonian}) splits
into sectors with different number of dimers (including possible overlaps
of dimers), and it is the sector with the minimal number of dimers (i.e.,
with non-overlapping dimers) which contains the RK ground state
(\ref{gs-interpolation}).

Furthermore, we can show that on the whole line $0<\mu<\infty$ the ground
state (\ref{gs-interpolation}) has a finite correlation length (at least,
on the commonly used lattices). It follows from the observation that the
equal-time correlations in the ground state (\ref{gs-interpolation}) are
exactly the same as in the classical fully frustrated Ising model without
the field defined on the same lattice at the dimensionless temperature
$T=1/\mu$. For quite a number of two-dimensional lattices the properties
of such models are known from the exact solutions (for a review, see,
e.g., Ref.\ \onlinecite{Liebmann}). In particular, the models on square
and triangular lattices are critical at $T=0$ and acquire a finite
correlation radius at an arbitrarily low temperature, whereas the models
on honeycomb, kagome and pentagon lattices have a finite correlation
radius already at zero temperature. In all these cases the correlation
radius continuously decreases with the increase in $T$ and at
$T\rightarrow\infty$ shrinks to zero.

We can also claim that this model (at $0<\mu<\infty$) has topological
order. Indeed, different topological sectors of the dimer representation
in terms of the Ising representation correspond to different boundary
conditions for spins, \cite{MS03} and the absence of translational
symmetry breaking in combination with a finite correlation radius
guarantees that these sectors are degenerate in the thermodynamic limit,
as expected in a topologically ordered state.

Note that for $\mu< \infty$ this topological order is of the Z$_2$ type
regardless of the lattice geometry (e.g., also for the square lattice).
This contrasts with the properties of the pure RK dimer model, which is of
the U(1) type on bipartite lattices, i.e., incorporates an infinite number
of topological sectors which can be characterized by integer ``winding
numbers.'' The reason for this reduction of symmetry is that the U(1)
conservation law for the dimer model on the square lattice breaks down to
Z$_2$ as soon as non-dimer states are allowed. A Z$_2$ topological order
implies also the existence of vortexlike Z$_2$ excitations (visons).
\cite{KRS} In fact, vison excitations in the Ising-spin language are
generated by the $\sigma^z_i$ operators.
\cite{Misguich-Serban-Pasquier,MS03}

While the ground state (\ref{gs-interpolation}) of our model is exactly
known for any lattice, the excitations are not. However, since the model
is of the RK type, then, as pointed out by Henley, the spectrum of the
lowest excitations can be efficiently computed with the {\em classical}
Monte Carlo method. \cite{Henley,Henley97} In principle, the spectra of
both vison and non-vison excitations can be computed by modeling a
classical stochastic walk in the space of configurations.
\cite{Ivanov04,LCA08}

\subsection{Generalized two-parameter model and its reduction to the FFIM}
\label{generalized-model}

A continuous interpolation between the FFIM with an arbitrary ratio
$\Gamma/J>0$ and a QDM with a variable parameter describing the relative
strength of different terms can be achieved by a slight modification of
the Hamiltonian (\ref{model1-hamiltonian}). It consists in ascribing
independent amplitudes to the potential and kinetic terms,
\begin{equation}                    \label{model-hamiltonian}
H(U,W) = U\sum_i \exp\left[ - \mu \sum_{j={\rm n.n.}(i)} \tau_{ij} \right]
-W\sum_i \sigma^x_i \, .
\end{equation}
Obviously, only the dimensionless ratio of $U$ and $W$ is of relevance for
the phase diagram.

This model reduces to an FFIM in the limit when $\mu$ is taken to $0$ at a
constant value of the product $U\mu$. In this limit, one can expand the
exponent in the first term of Eq.~(\ref{model-hamiltonian}) and keep only
the terms linear in $\mu$, because the coefficients in front of all higher
order terms vanish. The summation of contributions from different
plaquettes then immediately leads to the FFIM Hamiltonian
(\ref{ffim-hamiltonian}) with $J=2U\mu$ and $\Gamma=W$. This approach is
universal in the sense that it works for any lattice.

The reductions of the model (\ref{model-hamiltonian}) to dimer models are
more delicate and have to be discussed separately for different types of
lattices. Below we explicitly analyze the model (\ref{model-hamiltonian})
on the square (Sec. \ref{sec:square}) and honeycomb (Sec.
\ref{sec:triangular}) lattices, which correspond to dimer models on square
and triangular lattices, respectively.

\section{Square lattice}
\label{sec:square}

On the square lattice, quantum dimer models can be
obtained from the model (\ref{model-hamiltonian})
in two different limits.

\subsection{QDM at $\mu\to 0$}

One possible reduction to the standard RK model (\ref{rk-hamiltonian}) is
obtained from  (\ref{model-hamiltonian}) by  taking the limit $\mu\to 0$
in a way different from that described in Sec. \ref{generalized-model}. At
$0<\mu\ll 1$, the dimer states (in which each site belongs to only one
dimer) are separated from all other states (not satisfying this rule) by
the gap of the order of $U\mu$. Therefore, if $U$ goes to infinity faster
than $\mu$ goes to zero, the energies of all states with overlapping
dimers go to infinity and the Hilbert space of the system is reduced to
that of the dimer model. In particular, when $\mu\to 0$  at
$U\mu^2=\mbox{const}$, only the second-order terms in the expansion of the
exponential remain finite. One can show that their contribution to the
potential energies of different dimer states is proportional to the number
of square plaquettes populated by two parallel dimers, which means that
the potential energy acquires the form of the first term in
Eq.~(\ref{rk-hamiltonian}) with $v=4U\mu^2$.

When only the dimer states are allowed, the action of the kinetic term
from Eq.~(\ref{model-hamiltonian}) on them is reduced to the possibility
of dimer flipping [like in Eq.~(\ref{rk-hamiltonian})] with amplitude
$t=W$. Thus, when $\mu$ tends to zero at a finite value of $U\mu^2$, the
model (\ref{model-hamiltonian}) is reduced to the RK model
(\ref{rk-hamiltonian}) with $v=4U\mu^2$ and $t=W$.

\begin{figure}[b]
\includegraphics[width=0.8\hsize]{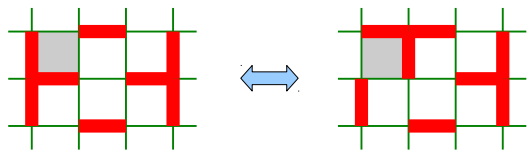}
\caption{(Color online) ``Star'' configurations of dimers on the square
lattice and their possible hopping.} \label{fig:stars-square}
\end{figure}

\subsection{QDM at $\mu\to\infty$}

Quite remarkably, another reduction to a dimer model is implemented in the
opposite limit $\mu\to\infty$ (at finite values of $U$ and $W$). In this
limit, potential energy vanishes for all plaquettes without dimers and for
plaquettes with one dimer, is equal to $U$ for plaquettes with two dimers
and is infinite for plaquettes with three or more dimers. This splits the
Hilbert space of the system into sectors with different number of dimers,
because all processes changing this number are prohibited. In particular,
the sector with the smallest possible number of dimers corresponds to the
standard situation when any site belongs to only one dimer. Within this
sector the Hamiltonian (\ref{model-hamiltonian})  is completely equivalent
to the RK Hamiltonian (\ref{rk-hamiltonian}) with $v=U$ and $t=W$.

Other sectors involve configurations with overlapping dimers (see Fig.\
\ref{fig:stars-square}) that we call below ``stars.'' Since each star adds
one extra dimer, these sectors can be labeled by the total number of stars
in the system. We can therefore refer to the model obtained in the $\mu
\to\infty$ limit as a ``star-dimer model.'' In terms of the height
representation for dimer coverings of the square lattice,
\cite{Levitov90,Henley97} each star corresponds to a screw dislocation,
whose sign depends on to which of the two sublattices this star belongs.

A relevant question in such a situation is whether the ``starless'' sector
is indeed the lowest-energy sector of the model. At $U=W$, one can show
that, while a single-star configuration has the same zero energy as the RK
ground state, starting from two stars, the ground-state energy becomes
positive. This is a consequence of a contact repulsive interaction between
stars: when stars are close to each other, some flips on plaquettes with
two dimers would produce plaquettes with three dimers and therefore are
prohibited. Accordingly, the kinetic energy cannot fully compensate the
potential one. Since on a bipartite lattice stars are ``charged'' (with
each star occupying three sites from one sublattice and only one from the
other), they can appear in a finite system only in pairs, and therefore
the one-star sector can be disregarded. We can then summarize our
conclusion that, at $U=W$, the starless sector has the lowest ground-state
energy.

Is is also possible to prove that the starless sector gives the
lowest-energy state at $U>W$. In this case, the staggered state, in which
each plaquette contains only one dimer and therefore no fluctuations are
possible, realizes the ground state of the model [the same nonflippable
state is known  \cite{RK} to be the ground state of the RK model
(\ref{rk-hamiltonian}) at $v>t$]. Indeed, since the full Hamiltonian may
be represented as
\begin{equation}                    \label{U>W-hamiltonian}
H(U,W)=H(W,W)+H(U-W,0)\, ,
\end{equation}
and the nonflippable state is a ground state of {\it both} $H(W,W)$
and $H(U-W,0)$ (at $U>W$), it must also be a ground state of $H(U,W)$
(without regard to a sector).

For $U<W$, the situation is more complicated: the sectors with stars may,
in principle, have lower energy than the ground state of the RK model.
While the latter is known to be crystal-ordered (with the plaquette
or columnar type of order \cite{LChR,Sachdev89,Syl06,RPM}),
it may be possible that in some interval
of $Y$ stars are energetically favorable (which would possibly modify or
destroy the crystal order). A simple perturbative study of stars doped
into the liquid at the RK point $Y=1$ suggests that stars are not
energetically favorable in the star-dimer model near the RK point.
However, for a rigorous justification of the absence of stars, a more
careful (possibly numerical) study is necessary, which takes into account
the presence of a crystalline order. We do not address this issue
in the present work, but leave it for future studies.

\begin{figure}
\includegraphics[width=0.8\hsize]{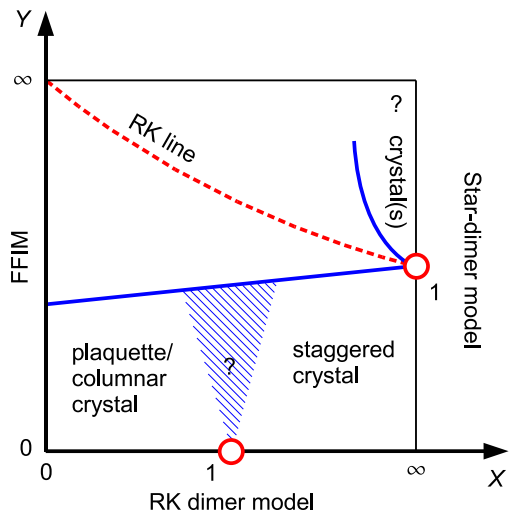}
\caption{(Color-online) A schematic representation of the phase diagram of
the two-parameter model (\ref{model-hamiltonian}) on the square lattice,
in the coordinates (\ref{coordinates-square}).} \label{fig:phase-square}
\end{figure}

\subsection{Phase diagram}

Constructing a full phase diagram of the model (\ref{model-hamiltonian})
is a challenging problem. The model depends on two dimensionless parameters
$\mu$ and $U/W$, and already the boundary of this parameter domain
contains several interesting models, including those discussed above
(FFIM, RK dimer model, and the star-dimer model).

For a better visualization of the discussed limiting cases, we
choose the coordinates for the phase diagram as follows:
\begin{equation}
X=\frac{U}{W}\sinh^2(2\mu)\,,~~~Y=\frac{W}{U}\coth(2\mu)\,.
\label{coordinates-square}
\end{equation}
Our proposal for the phase diagram in these coordinates is schematically
shown in  Fig.~\ref{fig:phase-square}. The vertical axis $X=0$ corresponds
to the FFIM with $\Gamma/J=Y$, and the horizontal axis $Y=0$ to the RK
model with $v/t=X$. The $X\to \infty$ limit corresponds to the star-dimer
model whose starless sector is given by the RK dimer model with
$t/v=Y$. At least for $0\leq Y\leq 1$, the true ground state of the model
(\ref{model-hamiltonian}) belongs to this sector.

In terms of the original variables $U$, $W$ and $\mu$, the point
$X=\infty$, $Y= 0$ corresponds to $W=0$ and, naturally, can be achieved at
any $\mu$ (which in terms of Fig.~\ref{fig:phase-square} corresponds to
approaching this point from any direction within the square). For any
$\mu$ the potential energy of the considered model is minimized in the
staggered states in which all plaquettes contain exactly one dimer,
$n_i\equiv 1$. Comparison of this energy with the potential energy of the
states with small numbers of plaquettes with $n_i\neq 1$ suggests that for
$W/U\ll \max(1,\mu)$ the quantum fluctuations on the background of a
staggered state are weak and cannot destroy it. In conjunction with
knowing that the staggered states are the ground states of the model both
at $X\geq 1$, $Y=0$ and at $X=\infty$, $Y\leq 1$, this allows one to
expect that in a finite region of the phase diagram the ground state
has the same type of ordering.

The ``RK line'' $U=W$ at which the ground state is given by
Eq.~(\ref{gs-interpolation}) corresponds to \makebox{$Y=1+1/\sqrt{X}$}.
Since at this line the system is in the RVB state with a finite
correlation radius, one expects that the RVB state occupies a finite
region on the phase diagram (in the vicinity of the RK line). This region
has to extend to the upper part of the axis $X=0$, which corresponds to
the translationally invariant disordered phase of the FFIM.

Below a certain critical value of $Y$, the FFIM is in an ordered phase
with a broken translational invariance. At $\Gamma/J=0$ (which corresponds
to the point \makebox{$X=Y=0$} in the lower left corner of our phase
diagram), this phase is connected to the crystal phase of the RK dimer
model at small values of $v/t$.

Furthermore, one can argue that the crystalline orders present in the RK
dimer model (i.e., at $Y=0$) extend to a finite region of the phase
diagram at $Y>0$. Indeed, at small $Y$, the potential energy of a
star is large, $E_{\rm st}\approx 4U\mu\approx 2t/Y$.
Therefore, at $Y\ll 1$, star configurations appear only as virtual pairs
and may be taken into account as perturbative corrections within the
quantum dimer model. In particular, the lowest order processes lead
to the renormalization of $v$ 
and to the appearance of an additional kinetic term
related to a cyclic flip of three dimers belonging to two neighboring
plaquettes. The amplitude of these corrections is proportional to $1/Y$,
and therefore their presence cannot destroy the crystalline phases of the
RK model in some region above the $Y=0$ line.

As revealed by numerical studies of the RK dimer model,
\cite{LChR,Syl06,RPM} at $0<v/t<1$ its phase diagram contains at least two
different crystal phases: the plaquette and columnar ones. Therefore, the
lower left part of our phase diagram must also contain {two or more
crystal phases.}

It follows from the analysis of Ref.~\onlinecite{FHMOS} that the
transition between the rightmost of these phases and the staggered crystal
{occupying} the lower right part of the phase diagram may occur along one
of the two scenarios: either as a first-order transition line terminating
at the RK point (\makebox{$X=1$}, $Y=0$) or as a devil's staircase
(complete or incomplete) of intermediate commensurate phases. {In either
case}, we do not expect any RVB region in the vicinity of the RK point.

Yet another crystal region is expected to exist at the vertical axis
$X=\infty$ above the RK point $Y=1$. As mentioned in the previous
subsection, a simple variational analysis of stars doped into the RK state
indicates that they should be energetically unfavorable just above the
$Y=1$ point. Therefore a plaquette crystal is expected in the dimer-star
model adjacent to this point.

A deviation from the $X=\infty$ axis leads to the mixing between the
different sectors of the star-dimer model. The difference from the region
just above the line $Y=0$ is that, in the vicinity of the $X=\infty$ axis,
the proper energy of a star is not high. However, at $X\gg 1$, the
amplitude of the formation of a pair of stars is low. In addition to that,
the presence of the crystalline order in the starless sector induces a
linear in distance attraction between the stars. It appears because, in
terms of the height representation, stars correspond to screw
dislocations. The combination of these factors leads to the confinement of
stars and does not allow them to destroy a crystalline order in some
vicinity of the $X=\infty$ axis (more precisely, of its part where the
star-dimer model is in the crystal phase). On the other hand, when
deviation from the $X=\infty$ axis takes place at $Y=1$, the confining
interaction between the stars induced by the crystalline order is absent
and one immediately gets into the RVB phase.

Further interesting phases may be possible in the upper right corner of
the phase diagram, with stars playing a role in the energetic balance. In
the present work, we do not explore this region of the phase diagram, but
leave this for future studies.

\vspace*{-2mm}
\section{Triangular QDM  / Honeycomb FFIM lattice}
\label{sec:triangular}

In this section, we consider the same model (\ref{model-hamiltonian}) in
the case of the triangular QDM lattice, which corresponds to the honeycomb
FFIM lattice. As on the square lattice, the FFIM limit is achieved in the
limit $\mu\rightarrow 0$ taken at $U\mu/W={\rm const}$, as is explained in
more detail in Sec. \ref{generalized-model}. The dimer limit is, however,
rather different from the square-lattice case. The reason for this
difference is that, on the triangular lattice, the constraint of
nonoverlapping dimers prohibits direct flips induced by operators
$\sigma^z_j$,  and the dynamics of dimers has to be mediated by virtual
flips via intermediate non-dimer states. Since the magnitude of the gap to
the nondimer states in the Hamiltonian (\ref{model-hamiltonian}) is of the
order of $U\sinh \mu$, the dimer model can be expected to be realized when
$W/(U\sinh \mu)  \to 0$. In this limit, the only terms which survive in
the effective Hamiltonian for the dimer model come from the second order
of the perturbation theory.

\begin{figure}[b]
\includegraphics[width=0.9\hsize]{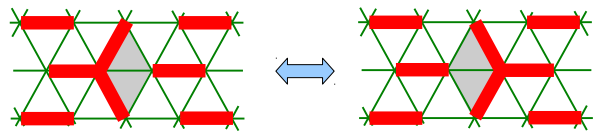}
\caption{(Color online) ``Star'' configurations of dimers on the
triangular lattice and their possible hopping.}
\label{fig:stars-triangular}
\end{figure}

Similarly to the ``star-dimer'' limit on the square lattice, the
peculiarity of the limit $W/(U\sinh \mu) \to 0$ on the triangular QDM /
honeycomb FFIM lattice is that the potential term in Hamiltonian
(\ref{model-hamiltonian}) does not impose the rigorous dimer constraint
but prohibits only having more than one dimer on every triangular
plaquette. In addition to the usual (non-overlapping) dimer coverings this
also permits the star-like overlaps of three dimers forming angles
$120^\circ$ with respect to each other (see
Fig.~\ref{fig:stars-triangular}). As a consequence, the effective model
obtained in the considered limit contains not only the dimer sector, but
also the sectors with such three-dimer stars. As in the star-dimer model
on a square lattice these sectors differ by the number of dimers: each
star adds one extra dimer.

\vspace*{-3mm}
\subsection{Rokhsar--Kivelson limit}

We first discuss the RK-point limit, which is realized when
$W/(U\sinh \mu)\to 0$ while $\mu\to \infty$. In this limit, the only terms
which survive in the effective Hamiltonian for the dimer model are related
to virtual flips on triangular plaquettes with one dimer. Analogous flips
on plaquettes without dimers can be neglected since they lead to
intermediate states with three dimers per plaquette (which for
$\mu\to\infty$  are infinitely higher in energy than the intermediate
states with two dimers per plaquette).  Then, in the dimer sector (without
stars), we arrive at the RK Hamiltonian with $v=t=W^2/(2U \sinh \mu)$.

In the case of sectors with stars, however, second-order perturbation
theory shows an imbalance between the potential and the kinetic terms.
Namely, for each star, there is a positive potential contribution of the
order $W^2/(U \sinh \mu)$, which is not compensated by a kinetic term. We
then conclude that configurations with stars are separated by a finite gap
from the RK dimer sector, the magnitude of the gap being of the order of
the coupling constant of the RK model.

\begin{figure}
\includegraphics[width=0.63\hsize]{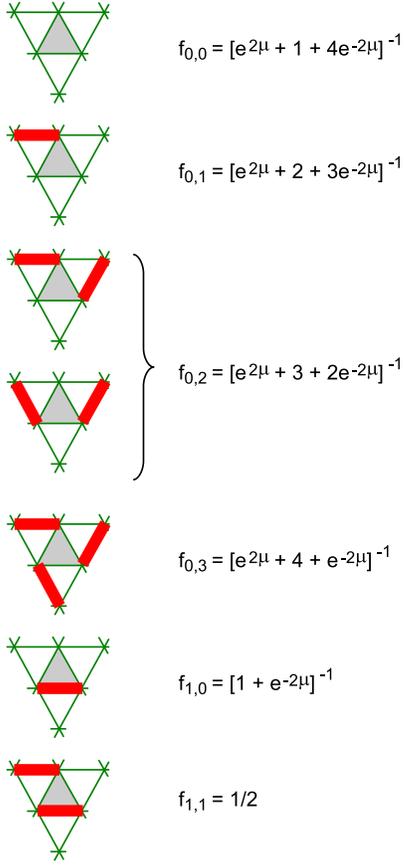}
\caption{(Color online) The coefficients $f_{n,\bar n}$ entering
Eq.~(\ref{E_i}) for different configurations of dimers on the plaquette
${\cal P}_i$ (shaded triangle in the middle) and the three neighboring
plaquettes.} \label{fig:weights-triangular}
\end{figure}

\vspace*{-3mm}

\subsection{QDM at finite $\mu$}

A more general QDM is obtained when the limit  $W/(U\sinh\mu) \to 0$ is
taken at a finite value of $\mu$. In this case, the effective kinetic term
is also of the RK type with $t=W^2/(2U \sinh \mu)$. The effective
potential energy becomes, however, more complicated, because at
$\mu<\infty$ the virtual flips on triangular plaquettes without dimers
also have to be taken into account.

In any configuration with only nonoverlapping dimers, a virtual flip on
the plaquette ${\cal P}_i$ produces a gain of the potential energy
\begin{equation}                                           \label{E_i}
E_i = - \frac{W^2}{2U\sinh\mu}\, f_{n_i,\bar n_i}(\mu)\,,
\end{equation}
where factors $f_{n_i,\bar n_i}(\mu)$ depend on $n_i$, the number of
dimers on ${\cal P}_i$, and $\bar n_i$, the total number of dimers
belonging to the three neighboring plaquettes of ${\cal P}_i$ but not
belonging to ${\cal P}_i$. Fig.~\ref{fig:weights-triangular} shows the
values of these factors for all possible configurations of dimers on
${\cal P}_i$ and the three neighboring plaquettes. It is not hard to
notice that $1/f_{n,\bar n}(\mu)$ depends linearly on both $n$ and $\bar
n$, which allows us to replace the six formulas shown in
Fig.~\ref{fig:weights-triangular} by a single expression,
\begin{equation}                                  \label{f-coefficients}
    f_{n,\bar n}(\mu)=[(1-n)e^{2\mu}+1+\bar n
    +(4-3n-\bar n)e^{-2\mu}]^{-1}.
\end{equation}

\begin{figure}[b]
\includegraphics[width=0.55\hsize]{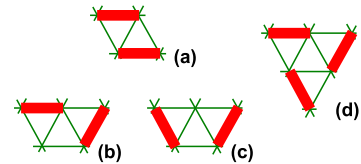}
\caption{(Color online) The potential energy given by the sum $\sum_i E_i$
may be reproduced by attributing energy $v$ to each pair of neighboring
dimers (a), energy $v_2$ to each pair of next-nearest-neighbor dimers [(b)
and (c)] and an additional energy $K$ to each three-dimer loop (d).}
\label{fig:interactions-triangular}
\end{figure}

The potential energy given by the sum $\sum_{i}^{}E_i$ may be
alternatively described in terms of local interactions between dimers. In
this description, in addition to $v$, the energy of the interaction of
nearest-neighbor dimers [see Fig.\ \ref{fig:interactions-triangular}(a)],
one needs to introduce two other coupling constants describing the
interaction of next-nearest-neighbor dimers. We denote the pairwise
interaction of such dimers as $v_2$ [see Figs.\
\ref{fig:interactions-triangular}(b) and
\ref{fig:interactions-triangular}(c)], whereas $K$ denotes an additional
three-body interaction in a loop of next-nearest-neighbor dimers [Fig.\
\ref{fig:interactions-triangular}(d)]. These definitions imply that the
dimer configurations shown in Figs.\ \ref{fig:interactions-triangular}(b)
and \ref{fig:interactions-triangular}(c) are ascribed the energy $v_2$,
whereas the configuration of Fig.\ \ref{fig:interactions-triangular}(d) is
ascribed the energy $3v_2+K$. After straightforward algebra one obtains
\begin{subequations}                                    \label{couplings}
\begin{eqnarray}
v/t &   = & \tanh\mu-2 (f_{0,0}- f_{0,1})\,,           \label{v/t}\\
v_2/t & = & - (f_{0,0} - 2 f_{0,1} + f_{0,2}) \,, \label{v2/t}\\
K/t   & = & f_{0,0} - 3 f_{0,1} + 3 f_{0,2} - f_{0,3} \,, \label{K/t}
\end{eqnarray}
\end{subequations}
where the argument of the functions $f_{0,\bar n}(\mu)$ is omitted.

\begin{figure}
\includegraphics[width=0.8\hsize]{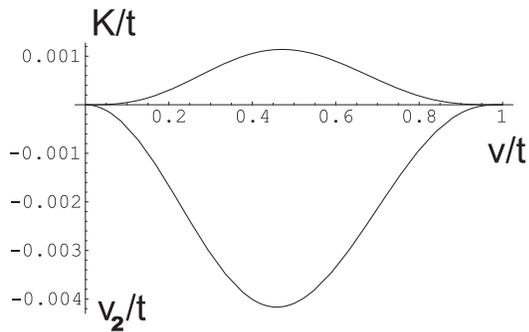}
\caption{$v_2$ and $K$ as functions of $v$.} \label{fig:couplings}
\end{figure}

In Eq.\ (\ref{v/t}), the relative magnitude of the second (negative) term
never exceeds $0.12$ and vanishes in the limit $\mu\to\infty$. The
functional dependence of $v/t$ on $\mu$ is monotonic, with $v/t$ ranging
from $0$ to $1$ as $\mu$ ranges from zero to infinity. Since the
magnitudes of coupling constants $v_2$ and $K$ are always numerically
small ($-0.0042<v_2/t< 0$ and $0<K/t<0.0012$, for any $\mu$ between zero
and infinity, see Fig.~\ref{fig:couplings}), the resulting QDM can
be understood as a small deformation of the RK Hamiltonian with $v/t$
given by Eq.\ (\ref{v/t}). At $\mu\to 0$ and at $\mu\to \infty$ both $v_2$
and $K$ tend to zero and one recovers the RK model (\ref{rk-hamiltonian})
with $v=0$ and $v=t$ respectively.

As for the star configurations, we have not analyzed their energies
between the two limits $\mu\to 0$ and $\mu\to\infty$. Since at
$\mu\to\infty$ their energy is positive and at $\mu\to 0$ it becomes
infinite, we conjecture that it remains positive also at intermediate
values of $\mu$, so that the dimer sector is always the lowest-energy
sector of the model. However, a rigorous verification of this property is
probably possible only numerically.

\subsection{Phase diagram}

\begin{figure}[b]
\includegraphics[width=0.8\hsize]{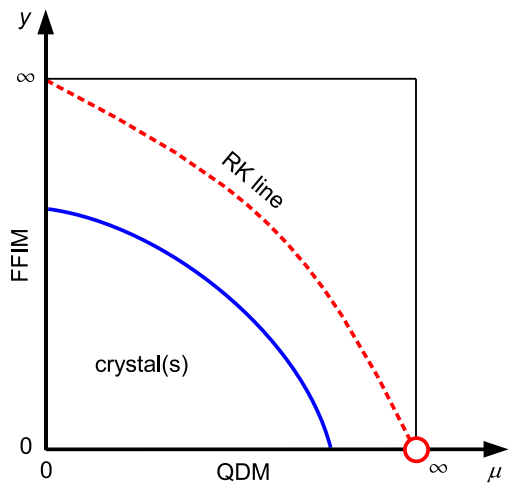}
\caption{(Color online) A schematic representation of the phase diagram of
the two-parameter model (\ref{model-hamiltonian}) on the triangular
lattice in the coordinates $\mu$ and $y=W/(2U\sinh\mu)$. }
\label{fig:phase-diagram-triangular}
\end{figure}

Based on the above consideration of the two limits, it is convenient to
plot the phase diagram of the model (\ref{model-hamiltonian}) on the
triangular lattice in the coordinates $\mu$ and
\begin{equation}
y=\frac{W}{2 U \sinh \mu}\,,
\end{equation}
see Fig.~\ref{fig:phase-diagram-triangular}. The vertical axis ($\mu=0$)
corresponds to the FFIM with $\Gamma/J = y$, and the horizontal axis
(\makebox{$y=0$}) to the QDM. This QDM differs slightly from the RK model
with $v/t\approx\tanh\mu$ due to the presence of the additional
interactions between the next-to-nearest dimers described by the coupling
constants $v_2$ and $K$ [see Eqs. (\ref{v2/t}) and (\ref{K/t})]. However
these interactions are always very weak and vanish when $\mu\to 0$ and
$\mu\to\infty$ when the model (\ref{model-hamiltonian}) becomes (in its
dimer sector) exactly equivalent to the RK model (\ref{rk-hamiltonian})
with $v=0$ and $v=t$ respectively.

At $v/t = 0.83(2)$ the RK dimer model on the triangular lattice
experiences a phase transition from the RVB to the crystal phase.
\cite{Ralko} A possible structure of this crystal phase has been discussed
not only in the context of the RK dimer model but also in the context of
the FFIM on the honeycomb lattice. The most likely crystal configuration
has a 12-site unit cell in the QDM formulation \cite{MS01prb,Ralko}
corresponding to a 24-site unit cell in the FFIM formulation.
\cite{Coletta} We expect that, similarly to the RK model, our limiting
dimer model obtained in the limit $y\to 0$ behaves in the same way. It
seems to us unlikely that the very weak additional interactions by which
this model differs from the RK model can lead to the formation of other
(more complicated) crystals. Since the crystal phase with the same
structure exists also in the FFIM at small enough values of $y$, it can be
expected to occupy a finite region in the lower left corner of the phase
diagram (see Fig.\ \ref{fig:phase-diagram-triangular}).

The RVB phase of the RK model spans the interval between $v/t \approx
0.83$ and $v/t=1$, which corresponds to $\mu > 1.22$, according to Eq.\
(\ref{v/t}). In this region, the next-nearest-neighbor interactions are
very weak. In particular, at the transition point $-v_2/t\sim 10^{-3} $
and $K/t\sim 10^{-4}$ with a further rapid decay at $v/t \to 1$, the
asymptotic behavior being \makebox{$v_2/t\approx-(1/4)(1-v/t)^3$} and
\makebox{$K/t\approx({3}/{8})(1-v/t)^4$}. This suggests  that the presence
of the additional interactions cannot induce a noticeable shift of the
position of phase transition in comparison with the RK model or have any
influence on the properties of the RVB phase.

The exactly solvable line is given in our coordinates by $y =1/(2\sinh
\mu)$. It follows from the exact solution of the classical
antiferromagnetic Ising model on the triangular lattice \cite{Stephenson}
that the ground state (\ref{gs-interpolation}) has a finite correlation
length everywhere on this line, and therefore we expect that this line
lies fully outside the crystal phase in our phase diagram. Note that the
staggered (non-flippable) phase of the RK dimer model, which appears at
$v/t>1$, is not present in our phase diagram, since this region of
parameters cannot be reached within our model. We can only span the region
between $v/t=0$ and $v/t=1$, with additional interactions of
next-nearest-neighbor dimers which are always too weak to stabilize the
staggered phase.

\vspace*{-4mm}
\section{Conclusion}
\label{sec:conclusion}

In this work, we have constructed a model interpolating between a quantum
dimer model and a fully frustrated Ising model in a transverse field and
studied its phase diagram for square and triangular (in terms of the dimer
representation) lattices. This interpolating model exhibits an RVB phase,
including a special ``RK line'' with an exactly solvable ground state.
This RK-type ground state generalizes the construction at the RK point of
the quantum dimer model, with its equal-time correlations given by the
classical fully frustrated Ising model (without a field) at a finite
temperature.

While the RK dimer model is a well-known ``prototype model''
of the RVB phase, our construction may be considered as its
generalization including ``star'' configurations. It can be
useful for studying the role of spin and charge degrees of
freedom in RVB states.

Indeed, a qualitative description of the RVB state in a spin system
contains elementary spin excitations (spinons) and vortexlike excitations
(visons). In application to high-temperature superconductivity, it has
been also suggested that doping RVB liquids gives rise to charge
excitations (vacancies or
holons).\cite{Read-Chakraborty,KRS,Senthil-Fisher} At the same time, the
RVB state in the pure dimer model only contains visons as elementary
excitations, with spinons and holons prohibited by the close-packing dimer
constraint. From this point of view, the model constructed in the present
work may serve as a prototype model of an RVB state including both visons
and holons (since star configurations carry extra charge).

Finally, we would like to note a certain similarity between the appearance
of RVB states on the square lattice close to the critical RK points in the
model of the present work (the point $X=\infty$, $Y=1$ in the phase
diagram in Fig.~\ref{fig:phase-square}) and in that of
Ref.~\onlinecite{Strubi}. In both situations, the initial RK state is
deformed into an RVB state by introducing a small concentration of charged
objects (stars in the present work and diagonal dimers in
Ref.~\onlinecite{Strubi}). One may note that, in both cases, the charged
defects cost initially zero energy and the deformation is achieved by
tuning the amplitude of their creation and annihilation. On the other
hand, if one starts with charged defects of infinite energy (which is the
case at the point $X=1$, $Y=0$ in the phase diagram in
Fig.~\ref{fig:phase-square}), then the critical dimer state on the
bipartite square lattice cannot be continuously deformed into an RVB state
(see also the discussion in Refs.~\onlinecite{FHMOS} and
\onlinecite{Vishwanath}).

There is also an interesting possibility that new phases may become
possible in our interpolating model, e.g., superfluid or supersolid ones.
We did not explore those possibilities in our analysis, but leave this
question for future studies.

\vspace*{-6mm} \acknowledgments \vspace*{-3mm}

The authors thank F.~Mila for useful discussions. S.E.K.\ acknowledges
support from RFBR Grant No.\ 09-02-01192a.

\end{document}